\documentclass[12pt,a4paper]{article}
\usepackage[latin1] {inputenc}
\usepackage[T1]{fontenc}
\usepackage[english]{babel}
\usepackage{amsmath}
\usepackage{amsfonts}
\usepackage{amssymb}
\begin{document}
\title{On the Casimir interaction between two smoothly deformed cylindrical surfaces}
\author{J.~D.~Melon~Fuksman$^{a,b}$
and
C.~D.~Fosco$^c$\\
\\
\\
{\normalsize\it $^a$Dipartimento di Fisica, Sapienza Universit\`a di Roma}\\
{\normalsize\it Piazzale Aldo Moro 5, 00185 Rome, Italy.}\\
\\
{\normalsize\it $^b$ICRANet}\\
{\normalsize\it Piazza della Repubblica 10, 65122 Pescara, Italy.}\\
\\
{\normalsize\it $^c$Centro At\'omico Bariloche and Instituto Balseiro}\\
{\normalsize\it Comisi\'on Nacional de Energ\'\i a At\'omica}\\
{\normalsize\it 8400 Bariloche, Argentina.}
}
\date{}
\maketitle
\begin{abstract}
\noindent We generalize the derivative expansion (DE) approach to the
interaction between almost-flat smooth surfaces, to the case of
surfaces which are optimally described in cylindrical coordinates. As
in the original form of the DE, the obtained method does not depend on the nature of the interaction. We apply our results to the study of the static, zero-temperature Casimir effect between two cylindrical surfaces, obtaining
approximate expressions which are reliable under the
assumption that the distance between those surfaces is always
much smaller than their local curvature radii. To obtain the zero-point energy, we apply known results about the thermal Casimir
effect for a planar geometry. To that effect, we relate the time
coordinate in the latter to the angular variable in the cylindrical
case, as well as the temperature to the radius of the cylinders. We study the dependence of the applicability of the DE on the kind of interaction, considering the particular cases where Dirichlet or Neumann conditions are applied to a scalar field.

\end{abstract}
\section{Introduction}\label{sec:intro}
The Casimir effect has been justly regarded as one of the most startling
{\em macroscopic\/} manifestations of  fluctuations, be them quantum
mechanical or thermal, of a field~\cite{booksCasimir}. 

To make predictions about the Casimir effect, typically involves evaluating
the influence of non-trivial boundary conditions on the vacuum (or thermal)
expectation values of the relevant observables. That task is, except when
highly symmetrical geometries are considered, rather involved.  One of the
main reasons for that is that those expectation values usually do not
satisfy a superposition principle, when regarded as functionals of the
boundary.  Thus, it is not possible, in general, to calculate the total
energy in the presence of a given boundary, by adding the contributions due
to each one of the possible pairs of surface elements into which the
boundary may be decomposed~\cite{CcapaTtira:2009ku}.  As a consequence,
rather few `universal' (i.e., applicable to an arbitrary surface)
properties of  the Casimir effect are known. 

The motivation to develop approximate methods to deal with rather general
geometries hardly needs to be emphasised. One of those methods, of much
wider applicability than the Casimir effect, is the so
called~\emph{Proximity Force Approximation} (PFA), originally introduced by
B.~Derjaguin in 1934~\cite{Derjaguin34}, within the context of the
interaction between interfaces.  This method has subsequently been applied
to several unrelated areas, like  nuclear physics~\cite{Blocki1,Blocki2},
Van der Waals interactions and, lately, the Casimir
effect~\cite{booksCasimir}, with varying degree of success.

In its most frequently used version, the PFA is applied to a setup
consisting of two interacting surfaces, $L$ and $R$, such that $L$ is
assumed to be a plane, and $R$, which (also by assumption) can be
represented by means of a single function, $\psi({\mathbf
x}_\shortparallel)$, the height of $R$ at each point ${\mathbf
x}_\shortparallel$ of $L$.  

Then, $E_{PFA}$, the PFA approximation to the interaction energy $E$
between the surfaces, is:
\begin{equation}\label{eq:1PFA}
E_{\rm PFA}\;\equiv\; \int d\sigma \,{\mathcal E}_{\shortparallel}\big[
\psi({\mathbf x}_\shortparallel) \big] \;,
\end{equation}
where $d\sigma$ is the area element at a point ${\mathbf
x}_\shortparallel$ on $L$, and ${\mathcal E}_\shortparallel(h)$ denotes the
energy per unit area for two parallel surfaces, i.e., for $\psi({\mathbf
x}_\shortparallel) \equiv h$, where $h$ is a constant~\footnote{The
approximation above can be generalized to two curved surfaces whenever they
may be both represented by two functions, $\psi_L$, $\psi_R$, which
measure the respective height about a common reference surface.}.

Until quite recently, there were no known {\em controlled\/} ways of
generalizing the PFA, so as to include shape-dependent corrections in 
an ordered perturbative expansion.  A step in that direction has been taken
with the introduction of the Derivative Expansion (DE)
~\cite{Foscoetal2011,bim1,bim2,Fosco:2012gp,Fosco:2012mn}, an approach that
leads to a modification of (\ref{eq:1PFA}) whereby the surface energy
density function includes derivatives of $\psi$, meant to account for a
dependence on the surface's local curvature. Successive terms in the
expansion have an increasing numbers of derivatives of $\psi$; the PFA
being reinterpreted as 
the zeroth (leading) order term in that expansion.  

This kind of approach is quite independent of the nature of the
interaction, what makes its potential range of applicability rather wide.
However, the implementation of the DE for surfaces that cannot be described
by using a single Monge patch is problematic, in part because of the
seemingly important role played by the Fourier transformation of the
distance function $\psi$, when written in Cartesian
coordinates~\cite{Fosco:2014rfa}, and also because of the different
topology of the manifolds.
This article presents an answer to that point, for the specific case of two
cylindrical surfaces. Besides implementing the DE, we also show that the
functions that define it may be
derived from known results obtained for a {\em planar\/} system at a finite
temperature~\cite{Fosco:2012mn}, by relating the periodicity of the
imaginary time to the one of the angular variable.  

Next-to-leading order corrections to the PFA have already been calculated
for particular cases of almost-cylindrical surfaces, for instance,
in~\cite{Lombardo:2008,CaveroPelaez:2008}. This kind of geometry is also interesting from the
experimental point of view, since it can be used to create configurations
that allow to measure lateral Casimir forces, as it is analyzed in these references. 

This paper is organized as follows: in Sect.~\ref{sec:DE}, after briefly
reviewing the DE in its usual, single Monge patch formulation in
~\ref{ssec:thede}, we present in~\ref{sec:DECyl} the analogous
construction for cylindrical surfaces.  Technical details of
the derivation are presented in an Appendix.  

In Sec.~\ref{sec:applications}, we apply the DE to the Casimir energy for a
quantum real scalar field satisfying either Dirichlet
or Neumann conditions. 
Finally, in Section~\ref{sec:concl}, we summarize our conclusions.  

\section{The Derivative Expansion}\label{sec:DE}
\subsection{Standard formularion}\label{ssec:thede}
We begin by reviewing the main features of the DE in its simplest setup:
two surfaces, $L$ and $R$, as the ones mentioned in the previous Section.
More specifically, we assume that a Cartesian coordinate system has been chosen
such that $L$ and $R$ occupy the regions (subsets of $\mathbb{R}^3$) given
by: \mbox{$s_L \equiv \{(x_1,x_2,0)\}$} and
\mbox{$s_R=\{(x_1,x_2,\psi({\mathbf x}_\shortparallel))\}$}, respectively.
Here, ${\mathbf x}_\shortparallel \equiv(x_1,x_2)$ and $\psi$ is a smooth
function of ${\mathbf x}_\shortparallel$.  Let $F[\psi]$ denote the
interaction energy between the two surfaces (not necessarily originated in
the Casimir effect) written as a functional of $\psi$. The DE yields an
approximation to $F$ as a series of local terms, ordered according
to their increasing number of derivatives of $\psi$. Up to the second
order~\footnote{Although in principle one could consider an arbitrary
number of orders, the number  of terms involved and the complexity
involved in their calculation grows rather fast.}:  
\begin{equation}\label{3aDE}
	F[\psi] \;=\; F_0[\psi] \,+\, F_2[\psi] \,+\,\ldots
\end{equation}
with 
\begin{equation}
F_0[\psi] \;=\; \int_{{\mathbf x}_\shortparallel}\, V(\psi({\mathbf
		x}_\shortparallel)) \;\;,\;\;\;
F_2[\psi] \;=\; \int_{{\mathbf x}_\shortparallel}\,Z(\psi({\mathbf x}_\shortparallel))
	|\nabla\psi|^2 \;
\end{equation}
(see, for example, \cite{Fosco:2014rfa}).

Once the functions $V$ and $Z$ are determined,  by any suitable method,
the previous equations may then be used to obtain approximate values for the
interaction energy between surfaces having {\em different} geometries,
namely, defined by different functions $\psi$.

One can see that the zeroth order term $F_0[\psi]$ above does reduce to the
PFA. Indeed, considering a (temporarily) finite integration area $S$, and a
constant $\psi({\mathbf x}_\shortparallel)=a$, all the terms but the first
one vanish. Hence, the function $V$ may be determined as follows:
\begin{equation}\label{3Va}
V(a) \;=\;\lim_{S\rightarrow\infty} \left(\frac{F[a]}{S}\right) \;.
\end{equation}
Thus, 
\begin{equation}
	F_0[\psi]=\int_{\mathbf{x}_{\vert\vert}} V(\psi({\mathbf
	x}_\shortparallel)) \;,
\end{equation}
which agrees, {\em mutatis mutandis\/} with the PFA (\ref{eq:1PFA}).

The next-to-leading-order (NTLO) term $F_2$, is in turn determined by the
$Z$ function, which may be obtained in an analogous way. For, example, 
one can evaluate $F[\psi]$ for $\psi({\mathbf x}_\shortparallel) = a +
\eta({\mathbf x}_\shortparallel)$, 
where $\eta$ is a function of ${\mathbf x}_\shortparallel$, whose mean value is $0$, 
while  $a$ is the average distance between the two surfaces. 
Expanding $F$ up to the second order in $\eta$, $Z$ may be extracted from the 
second order term in a momentum expansion of the Fourier transform of
$F$~\cite{Fosco:2014rfa}.

\subsection{DE for cylindrical surfaces}\label{sec:DECyl}
We present here the  conventions and results about the DE, when applied to
cylindrical surfaces, assuming their geometries may be naturally described
in terms of cylindrical coordinates $(\rho, \varphi, z)$. 
Details regarding the derivation of this result are presented in the Appendix. 

The geometry corresponds again to two surfaces, which we now
denote by $I$ and $O$. We retain the property that one of them is a coordinate 
surface, and the other can be defined by giving the distance of each one of 
its points to the first one. Indeed, we assume now $s_I$ to be a constant-$\rho$ coordinate surface,
namely,  a circular cylinder of radius $\rho = a$,  while $R$ occupies 
a region $s_O$ such that, for any given value of $\varphi$ and
$z$, its radius is determined by a single function $\psi$: $\rho =
\psi(\varphi,z)$. As in the previous subsection, we decompose $\psi$
into its average and its departure about it: $\psi(\varphi,z) \;=\; b
\,+\, \eta(\varphi,z)$.
 
The procedure outlined in the Appendix shows that the DE, up to the  
second order, is given by the expression:
\begin{equation}\label{eq:Cyl1}
F[\psi]\;=\; F_0[\psi]\;+\;F_2[\psi]\;+\; \ldots 
\end{equation}
where
\begin{equation}\label{eq:Cyl2}
	F_0[\psi] \;=\; \int_x b\,\mathcal{F}_0(b+\eta(x))
\end{equation}	
and
\begin{align}\label{eq:Cyl3}
F_2[\psi] \;=\;	\int_x\bigg\{ &
Z_1(\psi(x))\left(\partial_z\psi\right)^2+
Z_2(\psi(x))\left(\partial_\varphi\psi\right)^2 \nonumber\\
& + Z_{12}(\psi(x))\left(\partial_\varphi\psi\right)
\left(\partial_z\psi\right)
\bigg\} \;.
\end{align}
This result relies upon the same assumptions as the Cartesian case,  except for
an extra condition, namely, that $\psi$ and its derivatives are periodic functions
of the angular variable. Besides, we have kept a mixed term involving
derivatives with respect to the angle and $z$. The reason to keep that
term is that one might want to apply the approximation to systems where an
external field breaks, for example, the invariance under $\varphi \to
-\varphi$. One may imagine, for example, the existence of an external
magnetic field along $z$.
When there are sufficient symmetries, that term
will of course vanish.

\section{Applications}\label{sec:applications}
In this Section, we apply the DE for cylindrical surfaces to the
interaction energy resulting from the Casimir energy for a real scalar
field.  We work within the functional integral approach, in the imaginary time
formulation, where the spacetime metric becomes the identity matrix when
Cartesian coordinates $x=(x_0,x_1,x_2,x_3)$ are adopted, $x_0$ denoting the
Euclidean (imaginary) time. Spatial coordinates are denoted
collectively by ${\mathbf x}$.

The vacuum energy, which we shall denote by $E_0$, may be obtained as the
zero-temperature limit of the free energy (see, e.g., \cite{ZinnJustin:2002ru}), by means of the expression
\begin{equation}
E_0 = - \lim_{\beta\rightarrow\infty}  \beta^{-1} \log \mathcal{Z} \;,
\end{equation}
where ${\mathcal Z}$
denotes the canonical partition function
for a temperature $T =\beta^{-1}$ (natural units where $k_B=1$ have been 
adopted). This expression must not be mistaken for the energy of
the free vacuum, since boundary conditions may -and will- be
included in $\mathcal Z$.

To include such boundary conditions, we will use two
$\delta$-functionals:  $\delta_I$ and $\delta_O$, respectively. 
$\mathcal{Z}$, which will be a functional of $\psi$, can 
then be written as follows:
\begin{equation}\label{2Zcondeltas}
{\mathcal Z}[\psi] \;=\;\int
\mathcal{D}\phi\,\delta_I[\phi]\delta_O[\phi]\,e^{-\mathcal{S}_0[\phi]}
 \;, 
 \end{equation}
 where the integral is over $\phi$ configurations which are periodic in the
 time interval $[-\frac{\beta}{2} , \frac{\beta}{2}]$, and ${\mathcal
 S}_0[\phi]$ is the free Euclidean action. This shall be given by:
\begin{equation}
	{\mathcal S}_0[\phi] \;=\; \frac{1}{2} \int d^4x (\partial \phi)^2
	\;,
\end{equation}
where the $x_0$ integral runs from $-\frac{\beta}{2}$ to
$+\frac{\beta}{2}$. Similarly, we define $\mathcal Z_0$ as the
partition function when no boundary conditions are applied. 

\subsection{Scalar field with Dirichlet conditions}\label{S:C4_Dirichlet}
We first consider a real scalar field $\phi$ and approximately cylindrical 
surfaces, upon which Dirichlet conditions are imposed. 
The world-volumes swept by those surfaces will be parametrized as $s_I=\{
(x_0,\rho\cos\varphi, \rho\sin\varphi,z): \rho=a \}$ and
$s_O=\{ (x_0,\rho\cos\varphi, \rho\sin\varphi,z):
\rho=\psi(\varphi,z) \}$. Here,  $\varphi \in [-\pi , \pi)$, $x_0 \in
(-\infty, \infty)$, and $z \in (-\infty , \infty)$. 

We assume that $\psi(\varphi,z)=b+\eta(\varphi,z)$, with $b>a$,
and $\eta$  a differentiable function such that 
$\vert\eta(\varphi,z)\vert\ll b-a \;, \;\; \forall\,\,\varphi,z$. Following
the derivation in the Appendix, we choose $b$ so that 
\begin{equation}
\int_{-\pi}^{\pi}\mathrm{d}\varphi\int_{-\infty}^\infty\mathrm{d}z\, 
\eta(\varphi,z)=0,
\end{equation}
and hence the surface $s_O$ is, on average, a cylinder of radius $b$, and
the first order term in the functional expansion in powers of $\eta$, vanishes. 

To impose the condition $\phi_{s_I,s_O}=0$, we insert 
in $\mathcal{Z}[\psi]$ the functionals $\delta_I[\phi]$ and
$\delta_O[\phi]$, defined in terms of auxiliary fields $\xi_I(y)$ and
$\xi_O(y)$ as:
\begin{equation}
\delta_I[\phi]=\int\mathcal{D}\xi_I
\exp\left[i\int_{y}\xi_I(y)\phi(y)\delta(\rho-a)\right]
\end{equation}
\begin{equation}\label{4FuncDeltas}
\delta_O[\phi]=\int\mathcal{D}\xi_O
\exp\left[i\int_{y}\sqrt{g(y_{\vert\vert})}\,\xi_O(y)\phi(y)\frac{
\delta(\rho-\psi(y_{\vert\vert}))}{\rho}
\right],
\end{equation}
where $y\equiv(x_0,\rho,\varphi,z)$, 
$y_{\vert\vert}\equiv(x_0,\varphi,z)$, $\int_{y}\equiv
\int_{-\pi}^{\pi}\mathrm{d}\varphi
\int_{-\infty}^\infty\mathrm{d}z
\int_{-\infty}^\infty\mathrm{d}x_0
\int_0^\infty \rho \,\mathrm{d}\rho$ and $g(y_{\vert\vert})$ is the
determinant of the metric induced on $s_O$. 

Integrating out $\phi$, we see that:
\begin{equation}
\frac{\mathcal{Z}[\psi]}{\mathcal{Z}_0}
=\int\mathcal{D}\xi_I\mathcal{D}\xi_O
\exp\left[
-\frac{1}{2}\int_{y_{\vert\vert},y'_{\vert\vert}}
\xi_A(y_{\vert\vert})
T_{AB}(y_{\vert\vert},y'_{\vert\vert})
\xi_B(y'_{\vert\vert})
\right],
\end{equation}
where $A$ and $B$ may be $I$ or $O$, and 
$T_{AB}(y_{\vert\vert},y'_{\vert\vert})$ are the components of a matrix kernel 
$\mathbb{T}$, defined as 
$\mathbb{T}(y_{\vert\vert},y'_{\vert\vert})=\mathbb{M}(y_{\vert\vert}
)\mathbb{D}(y_{\vert\vert},y'_{\vert\vert})\mathbb{M}(y'_{\vert\vert})$, 
with
\begin{equation}
\mathbb{M}(y_{\vert\vert})=\begin{pmatrix}
a & 0 \\
0 & \sqrt{g(y_{\vert\vert})}
\end{pmatrix}=
\begin{pmatrix}
a & 0 \\
0 & \sqrt{(\partial_\phi\psi)^2+\psi^2(1+(\partial_z\psi)^2)}
\end{pmatrix},
\end{equation}
and 
$\mathbb{D}(y_{\vert\vert},y'_{\vert\vert})=\left(\begin{smallmatrix}
D_{II} & D_{IO} \\
D_{OI} & D_{OO}
\end{smallmatrix}\right)$. The latter are:
\begin{equation*}
D_{II}(y_{\vert\vert},y'_{\vert\vert})=\langle y_{\vert\vert},a| 
(-\partial^2)^{-1} | y_{\vert\vert}',a\rangle
\end{equation*}
\begin{equation*}
D_{IO}(y_{\vert\vert},y'_{\vert\vert})=\langle y_{\vert\vert},a | 
(-\partial^2)^{-1} | y_{\vert\vert}',\psi(\varphi',z')\rangle
\end{equation*}
\begin{equation*}
D_{OI}(y_{\vert\vert},y'_{\vert\vert})=\langle 
y_{\vert\vert},\psi(\varphi,z)| (-\partial^2)^{-1} | 
y_{\vert\vert}',a\rangle
\end{equation*}
\begin{equation}\label{4MatD}
D_{OO}(y_{\vert\vert},y'_{\vert\vert})=\langle 
y_{\vert\vert},\psi(\varphi,z)| (-\partial^2)^{-1} | 
y_{\vert\vert}',\psi(\varphi',z')\rangle
\end{equation}
where $\langle y_{\vert\vert},\rho| (-\partial^2)^{-1} | 
y_{\vert\vert}',\rho'\rangle$ is the (free) $\phi$-field propagator.

Thus, neglecting irrelevant contributions:
\begin{equation}\label{eq:sss1}
	E_0=\lim_{\beta\rightarrow\infty}\frac{1}{2\beta}{\rm 
Tr}\log\mathbb{T} \;.
\end{equation}
In the following subsections, we consider the different terms in the expansion
of (\ref{eq:sss1}) in powers of $\eta$ which are needed to construct the DE.

\subsubsection{$0^{th}$-order term}\label{S:C4_Orden0}
To this order, we need to take $\psi \equiv b$, and find the resulting
matrix elements $\mathbb{T}$. These may be obtained using the propagator
definition explicitly, writing the momenta $\mathbf{k}_{\vert\vert}=(k_1,k_2)$
in polar coordinates, and using the identities 
\begin{equation}\label{4IdBessel}
e^{i\,x\cos\alpha}=\sum_{m=-\infty}^\infty
i^mJ_m(x)\,e^{m\alpha}
\end{equation}
and
\begin{equation}\label{4IntJ2}
\int_0^\infty\mathrm{d}s\,
\frac{s}{\mathbf{k}_{\vert\vert}^2+s^2}J_n^2(s 
a)=I_n(\vert\mathbf{k}_{\vert\vert}\vert 
a)K_n(\vert\mathbf{k}_{\vert\vert}\vert a),
\end{equation}
which is valid for every $n\in\mathbb{Z}$. In these expressions, $J_n$ are Bessel functions of order $n$, 
while $I_n$ and $K_n$ denote modified Bessel functions. This leads to the 
result:
\begin{equation}\label{4TOrden0}
\mathbb{T}(y_{\vert\vert},y'_{\vert\vert})=\int_{\mathbf{k}_{\vert\vert}}
e^{i \mathbf{k}_{\vert\vert}
(\mathbf{y}_{\vert\vert}-\mathbf{y}'_{\vert\vert})}
\frac{1}{2\pi}
\sum_n e^{in(\varphi-\varphi')}\,
\widetilde{\mathbb{T}}(\mathbf{k}_{\vert\vert},n),
\end{equation}
with
\begin{equation}\label{4TtildeOrden0}
\widetilde{\mathbb{T}}(\mathbf{k}_{\vert\vert},n)=
\begin{pmatrix}
a^2\,
I_n(\vert\mathbf{k}_{\vert\vert}\vert 
a)K_n(\vert\mathbf{k}_{\vert\vert}\vert a) & 
a b\,
I_n(\vert\mathbf{k}_{\vert\vert}\vert 
a)K_n(\vert\mathbf{k}_{\vert\vert}\vert b)  \\
a b\,
I_n(\vert\mathbf{k}_{\vert\vert}\vert 
a)K_n(\vert\mathbf{k}_{\vert\vert}\vert b) & 
b^2\,
I_n(\vert\mathbf{k}_{\vert\vert}\vert 
b)K_n(\vert\mathbf{k}_{\vert\vert}\vert b)
\end{pmatrix} \;,
\end{equation}
and $\mathbf{y}_{\vert\vert} \equiv (x_0,z)$.

Thus, the  interaction energy per unit length $\mathcal{E}^l_0$ becomes:
\begin{equation}
\mathcal{E}^l_0=\lim_{\beta\rightarrow\infty,L\rightarrow\infty}
\frac{1}{2\beta L} {\rm Tr} \log \mathbb{T} =
\frac{1}{2} \int_{\mathbf{k}_{\vert\vert}}
\sum_n \log \det \widetilde{\mathbb{T}}(\mathbf{k}_{\vert\vert},n) \;.
\end{equation}
Evaluating the determinant, and discarding contributions which represent
self-energy terms (i.e., depending on each separate surface), we arrive to
the result:
\begin{equation}\label{4Orden0ECilindros}
\mathcal{E}^l_0=\frac{1}{2} \int_{\mathbf{k}_{\vert\vert}}
\sum_n \log\left[
1-\frac{I_n(\vert\mathbf{k}_{\vert\vert}\vert 
a)K_n(\vert\mathbf{k}_{\vert\vert}\vert 
b)}{I_n(\vert\mathbf{k}_{\vert\vert}\vert 
b)K_n(\vert\mathbf{k}_{\vert\vert}\vert a)}
\right],
\end{equation}
which is valid for any $a < b$.  This agrees with the known result for this 
case~\cite{EccentricExactSol}.

We know that the energy per unit area corresponding to the above result
should approach the analogous result for a couple of parallel planes when
the cylinders are sufficiently close to each other. Let us study this now, 
deriving at an intermediate step an approximate expression, which is
neither the result for cylinders nor for planes: it will correspond to
planes with a periodic coordinate, related to the angular variable.
When $d\equiv  b-a\ll a$, we can use the $n\rightarrow\infty$ 
approximations~\cite{watson,abram}: 
\begin{equation}
I_n(n\,z)\approx\sqrt{\frac{t}{2\pi n}}\,e^{n\xi(z)}
\end{equation}
and 
\begin{equation}\label{4AproxBessel}
K_n(n\,z)\approx\sqrt{\frac{\pi t}{2 n}}\,e^{-n\xi(z)},
\end{equation}
where $t=\frac{1}{\sqrt{1+z^2}}$ and 
$\xi(z)=\sqrt{1+z^2}+\log(\frac{z}{1+\sqrt{1+z^2}})$. 

Thus, when $d\ll a$, the ratio in (\ref{4Orden0ECilindros}) can be 
approximated as follows
\begin{align}\label{4InKnAprox}
\frac{I_n(\vert\mathbf{k}_{\vert\vert}\vert 
a)K_n(\vert\mathbf{k}_{\vert\vert}\vert 
b)}{I_n(\vert\mathbf{k}_{\vert\vert}\vert b)
K_n(\vert\mathbf{k}_{\vert\vert}\vert a)} &\approx
e^{2n[\xi(z_1)-\xi(z_2)]} \nonumber\\
& = e^{-2n\left(\sqrt{1+z_2^2}-\sqrt{1+z_1^2}\right)}e^{-2n\log\big(\frac{z_2}
{z_1}\frac{1+\sqrt{1+z_1^2}}{1+\sqrt{1+z_2^2}}\big)},
\end{align}
with $z_1=\vert\mathbf{k}_{\vert\vert}\vert a/n$ and 
$z_2=\vert\mathbf{k}_{\vert\vert}\vert b/n$. Next, expanding the 
exponents in (\ref{4InKnAprox}) for $d\ll a$, we have found that the
most accurate way to do so is to write the result in terms of $d$ and
$r=(b+a)/2$,  obtaining:
\begin{equation}
\frac{I_n(\vert\mathbf{k}_{\vert\vert}\vert 
a)K_n(\vert\mathbf{k}_{\vert\vert}\vert 
b)}{I_n(\vert\mathbf{k}_{\vert\vert}\vert 
b)K_n(\vert\mathbf{k}_{\vert\vert}\vert a)}\approx
e^{-2d\sqrt{(n/r)^2+\mathbf{k}_{\vert\vert} ^2}} \;.
\end{equation}
Therefore,
\begin{equation}\label{4E0lAprox}
\mathcal{E}^l_0\approx\frac{1}{2} \int_{\mathbf{k}_{\vert\vert}}
\sum_n \log\left(
1-e^{-2d\sqrt{(n/r)^2+\mathbf{k}_{\vert\vert} ^2}}
\right) \;,
\end{equation}
which is the intermediate expression mentioned above; indeed, it contains a
sum over a discrete `momentum', corresponding to the angular variable.
This expression, when divided by $2 \pi r$, yields the energy per unit area
$\mathcal{E}_0(r)$.  Moreover, it tends to the proper limit, i.e., to the parallel
planes result when $r\rightarrow\infty$:
\begin{equation}
\mathcal{E}_0(\infty)=\lim_{r\rightarrow\infty}\mathcal{E}_0(r)=
\frac{1}{2}\int_{k_{\vert\vert}}\log\left(
1-e^{-2d\vert k_{\vert\vert}\vert}
\right) \;.
\end{equation}
The way in which the limit is reached, may be studied by considering the
difference between these two magnitudes, 
\begin{equation}
\mathcal{E}_0(r)-\mathcal{E}_0(\infty)=2\,r\int_{-\infty}^\infty 
\frac{1}{2\pi}\int\mathrm{d}t\int_{(k_0,k_1)}\frac{\log\left(1-e^{-2d\sqrt{
k_0^2+k_1^2-(\epsilon+i\,t)^2}}\right)}{e^{2\pi r (\epsilon+i\,t)}-1} \;,
\end{equation}
(where we converted the series to an integral), which is not analytic at
$r\rightarrow\infty$, since  $e^{\alpha x}$ has an
essential singularity at $x\rightarrow\infty$. Namely, it is not possible to expand 
$\mathcal{E}_0(r)-\mathcal{E}_0(\infty)$ as a series in powers of $1/r$.

\subsubsection{$2^{nd}$ order}\label{S:C4_Orden2} 
To obtain the function $Z_2$ in the DE, we evaluate in this subsection
the second order term in the energy. We do this by applying the analogy
between the cylindrical geometry and a planar system
at a finite temperature, in the limit $d\ll r$. The approximation will be
checked in the concrete example for which the exact result is known,
namely, that of eccentric cylinders.

To perform this comparison, we consider the partition functions
$\mathcal{Z}^P[\psi]$ for approximately flat surfaces described in
Cartesian coordinates, and $\mathcal{Z}^C[\psi]$, for cylindrical
surfaces:
\begin{equation}\label{4_ZP}
\frac{\mathcal{Z}^P[\psi]}{\mathcal{Z}^P_0}=\int\mathcal{D}\xi_L\mathcal{D}
\xi_R
\exp\left[
-\frac{1}{2}\int_{x_{\vert\vert},x'_{\vert\vert}}
\xi_A(x_{\vert\vert})
K_{AB}(x_{\vert\vert},x'_{\vert\vert})
\xi_B(x'_{\vert\vert})
\right]
\end{equation}
\begin{equation}\label{4_ZC}
\frac{\mathcal{Z}^C[\psi]}{\mathcal{Z}^C_0}=\int\mathcal{D}\xi_I\mathcal{D}
\xi_O
\exp\left[
-\frac{1}{2}\int_{y_{\vert\vert},y'_{\vert\vert}}
\xi_A(y_{\vert\vert})
T_{AB}(y_{\vert\vert},y'_{\vert\vert})
\xi_B(y'_{\vert\vert})
\right].
\end{equation}
These expressions are evidently different; indeed,  even since the
components of $x_{\vert\vert}$ and $y_{\vert\vert}$ have different dimensions. 
To make the comparison less awkward, we  replace $\varphi$ by
$x_N=r\varphi\in[-\pi r,\pi r)$. This implies that, at least in the
$r\rightarrow\infty$ limit, the two partition functions should agree.

Performing that change of variables in (\ref{4_ZC}), we obtain an
additional $r^2$ factor, which leads to the conclusion that
$\mathcal{Z}^P[\psi]/\mathcal{Z}^P_0=\mathcal{Z}^C[\psi]/\mathcal{Z}^C_0$
is equivalent to the equality between the kernels
\begin{equation}\label{4LimEDM}
\frac{T_{AB}(y_{\vert\vert},y'_{\vert\vert})}{r^2}=K_{AB}(x_{\vert\vert},
x'_{\vert\vert}).
\end{equation}
When $r$ is much larger than $d$ but still finite, the integral of one of
the momenta in the calculation of $T_{AB}$ should be replaced by a sum over
discrete momenta, as it happened for the $0^{th}$ order, since one of the
coordinates is periodic. We have at our disposal the calculation for one
such system: two almost planar surfaces at a finite temperature
$T$. In that kind of system, the fields are periodic in the imaginary time:
$x_0\in[-1/2T,1/2T]$, where $T$ is the temperature.
Therefore, to use the results of such calculation, it is enough to replace $\beta$ by $2\pi r$.

The second order of $\Gamma_P \equiv -\log\mathcal{Z}_P$, which may be
extracted from \cite{Fosco:2012mn}, is:
\begin{equation}\label{6}
\Gamma_P^{(2)}[\psi]=\frac{1}{2\beta}\sum_n
\int_{\mathbf{k}_{\vert\vert}\equiv(k_1,k_2)}f^{(2)}(n,\mathbf{k}_{
\vert\vert})\,\vert\tilde{\eta}_P(n,\mathbf{k}_{\vert\vert})\vert^2,
\end{equation}
with
\begin{equation}\label{4etaP}
\tilde{\eta}_P(n,\mathbf{k}_{\vert\vert})=
\int_{(x_0,x_1,x_2)\equiv(x_0,\mathbf{x}_{\vert\vert})}\eta_P(x_0,\mathbf{x
}_{\vert\vert})\,e^{-i\, \mathbf{k}_{\vert\vert}.\mathbf{x}_{\vert\vert}} 
e^{-i\,(n/r)\, x_0}
\end{equation}
and
\begin{equation*}
f^{(2)}(n,\mathbf{k}_{\vert\vert})=
-\frac{1}{\pi r\,d^4}\sum_m\int_{\mathbf{p}_{\vert\vert}\equiv(p_1,p_2)}
\sqrt{(m d/r)^2+\mathbf{p}_{\vert\vert}^2}\sqrt{\left[(m+n) 
d/r\right]^2+\left( \mathbf{p}_{\vert\vert}
+\mathbf{l}_{\vert\vert}\right)^2}
\end{equation*}
\begin{equation}\label{f}
\times
\frac{1}{1-e^{-2 \sqrt{(m d/r)^2+\mathbf{p}_{\vert\vert}^2}}}
\,\frac{1}{e^{2 \sqrt{\left[(m+n) d/r\right]^2+\left( 
\mathbf{p}_{\vert\vert}
+\mathbf{l}_{\vert\vert}\right)^2}}-1},
\end{equation}
with $\mathbf{l}_{\vert\vert}=d\,\mathbf{k}_{\vert\vert}$.

Let us apply (\ref{S:C4_Orden2}) to a concrete example, that of 
two eccentric cylinders. This will allow us to find the explicit
form of the function $Z_2$ involved in the proposed DE.
We consider the external cylinder to be perturbed by a function
$\eta_C(\varphi)=\epsilon\cos \varphi$. In the limit $\epsilon\ll
d$, this describes two slightly eccentric cylinders, whose axes
are separated by a distance $\epsilon$. 

To obtain the interaction energy, we need
$\tilde{\eta}_P(n,\mathbf{k}_{\vert\vert})$. Setting
$\eta_C(\varphi)=\eta_P(x_N)$, we see that
\begin{equation}\label{010}
\vert\tilde{\eta}_P(n,\mathbf{k}_{\vert\vert})\vert^2=
(2\pi)^2\,r^2\,L^2\,\delta(\mathbf{k}_{\vert\vert})\,\pi^2\epsilon^2\,
(\delta_{n,1}
+\delta_{n,-1}) \;.
\end{equation}
Thus:
\begin{equation}\label{011}
\lim_{L\rightarrow\infty}\frac{\Gamma^{(2)}_P[\psi]}{L^2}=
\frac{\pi r}{4}\,\epsilon^2\,
[f^{(2)}(1,\mathbf{0})+f^{(2)}(-1,\mathbf{0})].
\end{equation}
Then we can use polar coordinates to perform the integral over 
$\mathbf{p}_{\vert\vert}$ in equation (\ref{f}). Defining 
$\rho=x/(\alpha-1)$, with $\alpha=b/a$, we may then write
the second order in $\eta$ of the interaction energy per
unit length, as 
\begin{equation*}
\mathcal{E}^{(2)l}_0[\psi]=-\frac{\epsilon^2}{8\pi a^4}
\sum_m\int_{0}^{\infty}\mathrm{d}\rho\,
\frac{\sqrt{[2m/(\alpha+1)]^2+\rho^2}}{1-e^{-2 (\alpha-1) 
\sqrt{[2m/(\alpha+1)]^2+\rho^2}}}
\end{equation*}
\begin{equation}\label{4E_L}
\times\left[\frac{\sqrt{[2(m+1)/(\alpha+1)]^2+ \rho^2}}{e^{2 (\alpha-1) 
\sqrt{[2(m+1)/(\alpha+1)]^2+\rho^2}}-1}+
\frac{\sqrt{[2(m-1)/(\alpha+1)]^2+ \rho^2}}{e^{2 (\alpha-1) 
\sqrt{[2(m-1)/(\alpha+1)]^2+\rho^2}}-1}
\right],
\end{equation}
which in the limit $\alpha\approx 1$ ($d\ll r$) reduces to:
\begin{equation*}
\mathcal{E}^{(2)l}_0[\psi]=-\frac{\epsilon^2}{8\pi a^4}
\sum_m\int_{0}^{\infty}\mathrm{d}\rho\,
\frac{\sqrt{m^2+\rho^2}}{1-e^{-2 (\alpha-1)\sqrt{m^2+\rho^2}}}
\end{equation*}
\begin{equation}\label{4E_L_aprox}
\times\left[\frac{\sqrt{(m+1)^2+ \rho^2}}{e^{2 (\alpha-1)\sqrt{(m+1)^2+ 
\rho^2}}-1}+\frac{\sqrt{(m-1)^2+ \rho^2}}{e^{2 (\alpha-1)\sqrt{(m-1)^2+ 
\rho^2}}-1}\right].
\end{equation}

The value of $\mathcal{E}^{(2)l}[\psi]$ thus obtained may be compared with
the second order term which follows from the exact result
in~\cite{EccentricExactSol}, where it is denoted as 
$E^{TM}/L$, since it corresponds to the transverse magnetic mode of the EM
field.
The second order of this exact solution is given by:
\begin{equation}\label{4E_TM_Ref}
\frac{E^{TM(2)}}{L}=-\frac{\epsilon^2}{4\pi a^4}\sum_n \int_0^\infty 
\mathrm{d}\rho\,\rho^3\frac{1}{1-\mathcal{D}_{n,n}^{TM,cc}}\left[
\mathcal{D}_n^{TM}+\frac{\mathcal{N}_n^{TM}}{1-\mathcal{D}_{n+1,n+1}^{TM,cc
}}
\right],
\end{equation}
where
\begin{equation}
\mathcal{D}_n^{TM}=\frac{\mathcal{D}_{n,n}^{TM,cc}}{2}+
\frac{I_n(\rho)}{4\,K_n(\rho)}\left[
\frac{K_{n-1}(\alpha\rho)}{I_{n-1}(\alpha\rho)}+
\frac{K_{n+1}(\alpha\rho)}{I_{n+1}(\alpha\rho)}
\right]
\end{equation}
\begin{equation}
\mathcal{N}_n^{TM}=
\frac{I_n(\rho)I_{n+1}(\rho)}{4\,K_n(\rho)\,K_{n+1}(\rho)}\left[
\frac{K_{n1}(\alpha\rho)}{I_{n1}(\alpha\rho)}+
\frac{K_{n+1}(\alpha\rho)}{I_{n+1}(\alpha\rho)}
\right]^2
\end{equation}
\begin{equation}
\mathcal{D}_{n,n}^{TM,cc}=\frac{I_n(\rho)\,K_n(\alpha\rho)}{K_n(\rho)\,
I_n(\alpha\rho)},
\end{equation}
and where $\epsilon$ is again the excentricity of the cylinders. To perform the comparison with our approximate expression, we first divided
them by $\epsilon^2/a^4$. 

Performing the sums and integrals numerically, we have found that while (\ref{4E_L})
is indeed a better approximation than (\ref{4E_L_aprox}) for $\alpha=1.1$
and bigger, they are quite similar for smaller values. 
In Table I, we show the comparison between (\ref{4E_L_aprox}) and
(\ref{4E_TM_Ref}). The error in the approximate expression decreases when 
$\alpha\rightarrow 1$, staying below $0.3\%$ when $\alpha<1.01$.

\begin{table}[htb]
\begin{center}
\begin{tabular}{|c|c|c|c|}
\hline
$\alpha$ & $\vert\mathcal{E}^{(2)l}_0\vert/(\epsilon^2/a^4)$ & $\vert 
E^{TM(2)}/L\vert/(\epsilon^2/a^4)$ & Error (\%) \\
\hline \hline
1.1 & 12,933.6 & 13,557.6 & 4.7\\ \hline
1.01 & 6.60 $10^8$ & 6.62 $10^8$ & 0.3\\ \hline
1.001 & 7.2034 $10^{12}$ & 7.2058 $10^{12}$ & 0.03\\ \hline
1.0001 & 7.21003 $10^{16}$ & 7.21012 $10^{16}$ & 0.001\\ \hline
1.00001 & 7.21010 $10^{20}$ & 7.21012 $10^{20}$ & 0.0003\\ \hline
\end{tabular}
\caption{Comparison, for different values of $\alpha$, between
	(\ref{4E_L_aprox}) and (\ref{4E_TM_Ref}).
The fourth column contains the error, defined as:
$100\times2\times(E^{TM(2)}/L-\mathcal{E}^{(2)l}_0)/(E^{TM(2)}/L+\mathcal{E}^
{(2)l}_0)$.}
\label{tabla:Tabla}
\end{center}
\end{table}

Finally, let us obtain the function $Z_2$ of the DE for this case,
based also on the example of
eccentric cylinders.  For $\psi=\psi(\varphi)$, this expansion reduces to:
\begin{equation}\label{4DDPhi}
E_0[\psi]=\int_0^{2\pi}\mathrm{d}\varphi
\left[
V(\psi(\varphi))+Z_2\left(\psi(\varphi)\right)\left(\frac{\partial\psi}{
\partial \varphi}\right)^2
\right].
\end{equation}
Now, setting $\psi(\varphi)=b+\eta(\varphi)$, with $\vert\eta\vert\ll b-a$, 
and expanding up to second order in $\eta$:
\begin{equation}\label{4_02}
E_0[b+\eta]\simeq\int_0^{2\pi}\mathrm{d}\varphi\left[
V(b)+V'(b)\,\eta(\varphi)+\frac{1}{2}\,V''(b)\,\eta^2(\varphi)
+Z_2(b)\left(\frac{\partial\eta}{\partial \varphi}\right)^2
\right].
\end{equation}

Setting now $\eta(\varphi)=\epsilon\cos\varphi$, we arrive to:
\begin{equation}
E_0[\psi]\simeq 2\pi V(b)+\frac{\pi\epsilon^2}{2}V''(b)
+\pi \epsilon^2 \,Z_2(b) \;.
\end{equation}
Hence we can extract the function $Z_2(b)$: 
\begin{equation}\label{4_04}
Z_2(b)=\frac{1}{\pi\epsilon^2} \left(
E_0[\psi]-2\pi V(b)-\frac{\pi\epsilon^2}{2}V''(b)
\right).
\end{equation}
At this point, it is useful to separate the total energy $E_0$ as a sum of 
its different orders in $\epsilon$. Doing so, we can see that the 
zeroth-order term equals $2\pi V(b)$. Hence, we are left with the 
following reduced expression:
\begin{equation}\label{4_05}
Z_2(b)=\frac{1}{\pi\epsilon^2}\left(
E_0^{(2)}[\psi]-\frac{\pi\epsilon^2}{2}V''(b)
\right),
\end{equation}
where $E_0^{(2)}[\psi]$ is the second order term (in $\epsilon$) of the energy.

At this time, we note that we can extract this term either from~\cite{EccentricExactSol},
or from the approximation (\ref{4E_L_aprox}).  We just need to evaluate the
second derivative of $V(b)$ with respect to $b$. Using our approximate
expression for concentric cylinders we obtain: 
\begin{equation}\label{4_06}
V(b)=\frac{L}{4\pi}\sum_{n}\int_{\mathbf{k}_{\vert\vert}}
\log \left(
1-e^{-2d\sqrt{(n/r)^2+\mathbf{k}_{\vert\vert}^2}}
\right) \;.
\end{equation}
Finally, we arrive to:
\begin{equation}
V''(b)=-\frac{L}{8\pi^2r^4}\sum_n\int_0^\infty\mathrm{d}\rho
\,\rho\, (\rho^2+n^2)\,\mathrm{cosech}^2\left((\alpha-1)\sqrt{\rho^2
+n^2}\right),
\end{equation}
with $\alpha=b/a$. 

$Z_2(b)$ may then be obtained by using our results for the energy.
\begin{equation*}
Z_2(b)=\frac{L}{4\pi^2}\sum_n\int_0^\infty\mathrm{d}\rho\,
\Bigg\{
\frac{\rho\, 
(\rho^2+n^2)}{4\,r^4}\,\mathrm{cosech}^2\left((\alpha-1)\sqrt{\rho^2
+n^2}\right)
\end{equation*}
\begin{equation}\label{4Z2_2}
-\frac{1}{2 a^4}\frac{\sqrt{(m+1)^2+ \rho^2}}{
1-e^{-2 (\alpha-1)\sqrt{m^2+\rho^2}}}
\left[\frac{\sqrt{(m+1)^2+ \rho^2}}{e^{2 (\alpha-1)\sqrt{(m+1)^2+ 
\rho^2}}-1}+\frac{\sqrt{(m-1)^2+ \rho^2}}{e^{2 (\alpha-1)\sqrt{(m-1)^2+ 
\rho^2}}-1}\right]
\Bigg\}.
\end{equation}
To sum up, equations (\ref{4DDPhi}), (\ref{4_06}) and (\ref{4Z2_2})
determine the  second order DE for the Dirichlet case.

\subsection{Scalar field: Neumann conditions}\label{S:C4_Neumann}

The same calculations can be performed when Neumann conditions are
imposed.  For this purpose, we choose the following boundary conditions:
\begin{align}
\partial_\rho\phi(y)\vert_{s_I}&=0 \\
\partial_n \phi(y)\vert_{s_O}&=0,
\end{align}
with $\partial_n\equiv n^\mu\partial_\mu$ and
$\partial_\mu\equiv\frac{\partial}{\partial x^\mu}$, where
$n^\mu(y_{\vert\vert})$ is a unit vector perpendicular to $s_O$, and
$x^{\mu}$ are usual Cartesian coordinates: 
\begin{equation}
n^\mu(y_{\vert\vert})=\frac{N^\mu(y_{\vert\vert})}{\vert 
N^\mu(y_{\vert\vert}) \vert},
\end{equation}
with
\begin{equation}
N^\mu(y_{\vert\vert})=
  \begin{cases} 
      \hfill  0    	\hfill & \text{ ($\mu=0$) } \\
      \hfill  \partial_\varphi\psi \sin\varphi +\psi \cos\varphi 	\hfill 
& \text{ ($\mu=1$)} \\
	  \hfill  -\partial_\varphi\psi \cos\varphi +\psi \sin\varphi	\hfill 
& \text{ ($\mu=2$) } \\
	   \hfill  -\psi\,\partial_z\psi	\hfill & \text{ ($\mu=3$) } \\
  \end{cases}.
\end{equation}

Again, we can include the boundary condition using functionals $\delta_I[\phi]$ and 
$\delta_O[\phi]$:
\begin{equation*}
\delta_I[\phi]=\int\mathcal{D}\xi_I
\exp\left[i\int_{y}\xi_I(y)\,\delta(\rho-a)\,\partial_{\rho}\phi(y)\right]
\end{equation*}
\begin{equation}\label{4FuncDeltasN}
\delta_O[\phi]=\int\mathcal{D}\xi_O
\exp\left[i\int_{y}\sqrt{g(y_{\vert\vert})}\,\xi_O(y)\,\frac{
\delta(\rho-\psi(y_{\vert\vert}))}{\rho}\,\partial_n\phi(y)
\right].
\end{equation}
Following analogous steps to those in Section~\ref{S:C4_Dirichlet}, 
$\mathcal{Z}$ may be written in a familiar form:
\begin{equation}\label{2_ZNeumannSinPhi}
\frac{\mathcal{Z}[\psi]}{\mathcal{Z}_0}=\int\mathcal{D}\xi_I\mathcal{D}
\xi_O
\exp\left[
-\frac{1}{2}\int_{y_{\vert\vert},y'_{\vert\vert}}
\xi_A(y_{\vert\vert})
N_{AB}(y_{\vert\vert},y'_{\vert\vert})
\xi_B(y'_{\vert\vert})
\right],
\end{equation}
where
\begin{align}\label{4NLL}
N_{II}(y_{\vert\vert},y'_{\vert\vert})&=a^2 \left[
{\partial_\rho}{\partial'_{\rho}}
\langle y \vert
(-\partial^2)^{-1}
\vert y' \rangle
\right]_{\rho=\rho'=a}\\
N_{IO}(y_{\vert\vert},y'_{\vert\vert})&=a\left[
{\partial_\rho}\partial'_N
\langle y \vert
(-\partial^2)^{-1}
\vert y' \rangle
\right]_{\rho=a,\,\rho'=\psi(y'_{\vert\vert})}\\
N_{OI}(y_{\vert\vert},y'_{\vert\vert})&=a \left[
\partial_N{\partial'_\rho}
\langle y \vert
(-\partial^2)^{-1}
\vert y' \rangle
\right]_{\rho=\psi(y_{\vert\vert}),\,\rho'=a}\\\label{4NRR}
N_{OO}(y_{\vert\vert},y'_{\vert\vert})&=\left[
\partial_N\partial'_N
\langle y \vert
(-\partial^2)^{-1}
\vert y' \rangle
\right]_{\rho=\psi(y_{\vert\vert}),\,\rho'=\psi(y'_{\vert\vert})},
\end{align}
with $\partial'_\rho\equiv\frac{\partial}{\partial\rho'}$ and 
$\partial'_N\equiv N^\mu(y'_{\vert\vert})\partial_{\mu}$. As before, this 
allows us to calculate the first orders in $\eta$ of the interaction 
energy.

\subsubsection{Order 0 in $\eta$}\label{S:C4_Orden0Neumann}
We start again with the order 0 in $\eta$. Following similar steps as 
before, we obtain that the matrix 
$\mathbb{N}(y_{\vert\vert},y'_{\vert\vert})$ may be written as:
\begin{equation}
\mathbb{N}(y_{\vert\vert},y'_{\vert\vert})=\int_{\mathbf{k}_{\vert\vert}}e^
{i\mathbf{k}_{\vert\vert}.(\mathbf{y}_{\vert\vert}-\mathbf{y}_{\vert\vert})
}\frac{1}{2\pi}\sum_n 
e^{in(\varphi-\varphi')}\,\widetilde{\mathbb{N}}(n,\mathbf{k}_{\vert\vert}),
\end{equation}
where
\begin{equation}
\widetilde{\mathbb{N}}(n,\mathbf{k}_{\vert\vert})=\mathbf{k}^2_{\vert\vert}
\begin{pmatrix} 
a^2\, I_n'(\vert \mathbf{k}_{\vert\vert}\vert a)K_n'(\vert 
\mathbf{k}_{\vert\vert}\vert a) & 
ab \,I_n'(\vert \mathbf{k}_{\vert\vert}\vert a)K_n'(\vert 
\mathbf{k}_{\vert\vert}\vert b) \\
ab \,I_n'(\vert \mathbf{k}_{\vert\vert}\vert a)K_n'(\vert 
\mathbf{k}_{\vert\vert}\vert b) &
b^2 \,I_n'(\vert \mathbf{k}_{\vert\vert}\vert b)K_n'(\vert 
\mathbf{k}_{\vert\vert}\vert b)
\end{pmatrix}.
\end{equation}
This matrix leads to the interaction energy per 
unit area:
\begin{equation}
\mathcal{E}_0^{(0)}=\frac{1}{4\pi r} \int_{\mathbf{k}_{\vert\vert}}\sum_n\, 
\log \det \widetilde{\mathbb{N}}(n,\vert \mathbf{k}_{\vert\vert}\vert)=
\frac{1}{4\pi r}\int_{\mathbf{k}_{\vert\vert}}\sum_n\, \log \left[
1-\frac{I_n'(\vert \mathbf{k}_{\vert\vert}\vert a)K_n'(\vert 
\mathbf{k}_{\vert\vert}\vert b)}{I_n'(\vert \mathbf{k}_{\vert\vert}\vert 
b)K_n'(\vert \mathbf{k}_{\vert\vert}\vert a)}
\right] \;,
\end{equation}
which coincides with the exact solution for concentric cylinders, 
computed in \cite{EccentricExactSol}.

On the other hand, one again expects the matrices $\mathbb{U}$ 
and $\mathbb{N}$ to satisfy an analogous relation to (\ref{4LimEDM}) in the 
limit $d\ll r$, i.e.,
\begin{equation}\label{4LimEDMNeumann}
\frac{\mathbb{N}(y_{\vert\vert},y'_{\vert\vert})}{r^2}\approx 
\mathbb{U}(x_{\vert\vert},x'_{\vert\vert}),
\end{equation}
where $\mathbb{U}$ is the equivalent to the matrix $\mathbb{K}$ in 
(\ref{4_ZP}), in the case where Neumann conditions are imposed. At order 0 
in $\eta$, this relation can be proved approximating the Bessel functions 
as in (\ref{4AproxBessel}), which gives:
\begin{equation}
\mathbb{N}^{(0)}(y_{\vert\vert},y'_{\vert\vert})\approx
-\frac{r}{4\pi}\int_{\mathbf{k}_{\vert\vert}}\sum_n
e^{i k_{\vert\vert}(y_{\vert\vert}-y'_{\vert\vert})}\, \vert 
k_{\vert\vert} \vert
\begin{pmatrix}
1 & e^{-d\vert k_{\vert\vert} \vert}\\
e^{-d\vert k_{\vert\vert} \vert} & 1 \\
\end{pmatrix},
\end{equation}
where, as before, 
$k_{\vert\vert}\equiv(\omega_n,\mathbf{k}_{\vert\vert})$, with 
$\omega_n=n/r$. Finally, we can use this to calculate the interaction 
energy per unit length:
\begin{equation}\label{4El0Neumannnn}
\mathcal{E}^l_0\approx\frac{1}{2}\int_{\mathbf{k}_{\vert\vert}}\sum_n \log
\left( 1-e^{-2d\sqrt{(n/r)^2+\mathbf{k}^2_{\vert\vert}}} \right).
\end{equation}
As it happens when Dirichlet conditions are imposed, if we divide this 
expression by $2\pi r$ and take the limit $r\rightarrow\infty$, we again 
obtain the energy density per unit area between parallel planes. On the 
other hand, equation (\ref{4El0Neumannnn}) leads to the same value of 
$\mathcal{E}_0^l$ obtained with Dirichlet conditions, in the limit $d\ll 
r$. Then, in that limit, the energy per unit length of the electromagnetic 
field coupled to perfect conductors shaped as $s_I$ and $s_O$ must be the 
following:
\begin{equation}
\mathcal{E}^{l(EM)}_0=
\mathcal{E}^{l(\text{Dirichlet})}_0+\mathcal{E}^{l(\text{Neumann})}_0
=2\,\mathcal{E}^{l(\text{Dirichlet})}_0
\approx\int_{\mathbf{k}_{\vert\vert}}\sum_n \log
\left( 1-e^{-2d\sqrt{(n/r)^2+\mathbf{k}^2_{\vert\vert}}} \right),
\end{equation}
which coincides with the limit $d\ll r$ of the exact solution (see
\cite{EccentricExactSol}). 

\subsubsection{Order 2 in $\eta$}\label{S:C4_Orden2Neumann}
We consider here the second order term. We shall see that, when Neumann
boundary conditions are imposed on $s_I$ and $s_O$, depending
on the variables upon which $\eta$ depends, the energy can have non
analytic properties that may render the
DE not applicable in certain cases. 

The second order term from $\Gamma_P[\psi]$,
calculated in~\cite{Fosco:2012mn}, is:
\begin{equation}\label{4_6_N}
\Gamma_P^{(2)}[\psi]=\frac{1}{2\beta}\sum_n
\int_{\mathbf{k}_{\vert\vert}}g^{(2)}(n,\mathbf{k}_{\vert\vert})\,
\vert\tilde{\eta}_P(n,\mathbf{k}_{\vert\vert})\vert^2,
\end{equation}
with
\begin{equation*}
g^{(2)}(n,\mathbf{k}_{\vert\vert})=
-\frac{1}{\pi r\,d^4}\sum_m\int_{\mathbf{p}_{\vert\vert}}
\frac{[m(m+n)(d/r)^2+\mathbf{p}_{\vert\vert}.(\mathbf{p}_{\vert\vert}
+\mathbf{l}_{\vert\vert})]^2}{\sqrt{(m 
d/r)^2+\mathbf{p}_{\vert\vert}^2}\sqrt{\left[(m+n) d/r\right]^2+\left( 
\mathbf{p}_{\vert\vert}
+\mathbf{l}_{\vert\vert}\right)^2}}
\end{equation*}
\begin{equation}\label{4g_N}
\times
\frac{1}{1-e^{-2\sqrt{(m d/r)^2+\mathbf{p}_{\vert\vert}^2}}}
\,\frac{1}{e^{2\sqrt{\left[(m+n) d/r\right]^2+\left( 
\mathbf{p}_{\vert\vert}
+\mathbf{l}_{\vert\vert}\right)^2}}-1},
\end{equation}
where $\mathbf{l}_{\vert\vert}=d\,\mathbf{k}_{\vert\vert}$, and $\beta=2\pi
r$ as before. 

The expansion of $g^{(2)}$ close to zero momentum can be used to obtain the
different orders in derivatives of $\eta$, which is not possible if
$g^{(2)}(n,\mathbf{k}_{\vert\vert})$ is not analytic in a neighbourhood of
$(n,\mathbf{k}_{\vert\vert})=(0,\mathbf{0})$. One way to verify this kind
of issue is to study the behaviour of the function
$g^{(2)}(0,\mathbf{k}_{\vert\vert})$ around $\mathbf{k}_{\vert\vert}=0$.
Examining equation (\ref{4g_N}), we can see that the terms with $m\neq 0$
will be analytic, since they are integrals of quotients of analitic
integrable functions that do not vanish. We still have to see the term with
$m=0$, for which we define $g(\mathbf{k}_{\vert\vert})$ as the term with
$m=0$ in (\ref{4g_N}) when $n=0$. Namely,
\begin{equation}
g(\mathbf{k}_{\vert\vert})=-\frac{1}{\pi 
r\,d^4}\int_{\mathbf{p}_{\vert\vert}}
\frac{[\mathbf{p}_{\vert\vert}.(\mathbf{p}_{\vert\vert}+\mathbf{l}_{
\vert\vert})]^2}{\vert \mathbf{p}_{\vert\vert} \vert \vert 
\mathbf{p}_{\vert\vert}+\mathbf{l}_{\vert\vert} \vert}
\frac{1}{1-e^{-2 \vert \mathbf{p}_{\vert\vert} \vert}}
\,\frac{1}{e^{2 \vert \mathbf{p}_{\vert\vert}+\mathbf{l}_{\vert\vert} 
\vert}-1}.
\end{equation}
A long calculation that involves dimensional regularization proves that, 
close to $\mathbf{k}_{\vert\vert}=\mathbf{0}$, this function behaves as
\begin{equation}
g(\mathbf{k}_{\vert\vert})\approx g(\mathbf{0})
-\frac{1}{32\pi^2\,r}\,\frac{\mathbf{k}^2_{\vert\vert}}{d^2}
\log\left(\mathbf{k}^2_{\vert\vert}\, d^2\right)
+
\mathcal{O}(\mathbf{k}_{\vert\vert}^2),
\end{equation}
where the term of order $\mathbf{k}^2_{\vert\vert}$ is finite.

Replacing it in equation (\ref{4_6_N}), a term proportional to
$\mathbf{k}^2_{\vert\vert}\log(\mathbf{k}^2_{\vert\vert}\, d^2)$ would give
rise to contributions to the energy proportional to
\begin{equation}
\int_{\mathbf{x}_{\vert\vert}}\int_{\varphi,z}\eta(\varphi,z)\,
\partial^2_z \log\left(-d^2\partial_z^2\right)\eta(\varphi,z),
\end{equation}
and therefore the proposed DE would not be applicable to this case. If
$\eta$ does not depend on $z$, however, $\vert
\tilde{\eta}_P(n,\mathbf{k}_{\vert\vert})\vert^2$ results to be
proportional to $\delta(\mathbf{k}_{\vert\vert})$, which nullifies the
contribution of terms such as
$\mathbf{k}^2_{\vert\vert}\log(\mathbf{k}^2_{\vert\vert}\, d^2)$ whose
limit as $\mathbf{k}_{\vert\vert}\rightarrow\mathbf{0}$ is $0$.
Consequently, the applicability of the DE depends in this case on the
analyticity of $g^{(2)}(n,\mathbf{0})$ as a function of $n$. On the other
hand, these problems with $\mathbf{k}_{\vert\vert}$ do not appear when
Dirichlet conditions are fixed, since in that case the zero-momentum 
expansion of the function equivalent to $g(\mathbf{k}_{\vert\vert})$ has
only the $\mathcal{O}(\mathbf{k}_{\vert\vert}^2)$ term, apart from the
constant one.

\section{Conclusions}\label{sec:concl}
We have constructed a version of the DE which is suitable for application
to cylindrical surfaces, and for a rather general interaction.
That expansion has then been applied to the Casimir effect at zero
temperature, for a real scalar field satisfying either Dirichlet or Neumann
conditions on two surfaces. We have shown how, in the limit where the DE 
yields approximate results, one can determine the functions appearing in
the DE approximation just from the knowledge of results for planar
surfaces at finite temperature. The role of the temperature is here of
course rather fictitious, since it is used (via the Matsubara formalism) 
to have a periodic coordinate.
We have checked numerically the intuitive idea that, when two cylindrical
surfaces are very close in comparison with the curvature radius, the
predictions coming from exact results are essentially the same as the ones
coming from planes with a periodic coordinate. We may say that, at least 
to the second order, the DE is sensitive to the topology (periodicity) of
the system, albeit not to the detailed geometry (the metric tensor).

In the Neumann case, the same known non-analyticity found at finite
temperature for planes arises. However, one can also show explicitly
that, if the  surfaces are translation invariant in $z$, the
non-analyticity disappears from the final expression.
\section*{Acknowledgements}
We are thankful to F.D. Mazzitelli for useful discussions that led to the
improvement of this work.

We acknowledge financial support from ANPCyT, CONICET, and UNCuyo
(Argentina).  We are especially grateful to Instituto Balseiro, where most
of this work has been done.

\newpage
\section*{Appendix: Derivation of the DE for two cylindrical surfaces}
We derive here the DE shown in Sect.~\ref{sec:DECyl} for cylindrical surfaces
(equations~(\ref{eq:Cyl1}) to~(\ref{eq:Cyl3})). For this, we consider two
surfaces: $I$, which corresponds to $\rho = a$, and $O$ to
$\psi(\varphi,z)=b+\eta(\varphi,z)$, with $b > a$. 

As in the case of Cartesian coordinates, we begin by assuming that the  functional 
$F[\psi]$, that represents the interaction energy, may be expanded
in a functional Taylor series:
\begin{equation}\label{3ExpF}
F[\psi] \;=\;\sum_{n\ge 0}\frac{1}{n!}
\int_{x_1,\ldots,x_n}\Gamma^{(n)}(x_1,\ldots,x_n)\,
\eta(x_1)\ldots\eta(x_n) \;,
\end{equation}
where we have used a shorthand notation $x \equiv (\varphi, z)$ for the 
integration variables. Namely,
\begin{equation}
\int_{x_1,\ldots,x_n} \ldots \; \equiv \; 
\int_{-\infty}^\infty\mathrm{d}z_1 \int_{-\pi}^\pi\mathrm{d}\varphi_1
\ldots 
\int_{-\infty}^\infty\mathrm{d}z_n \int_{-\pi}^\pi\mathrm{d}\varphi_n
\ldots
\end{equation}
Since we want to deal with smooth functions, $\eta$ (and therefore $\psi$) must 
be $2\pi$-periodic in its angular argument $\varphi$. 

The functional derivatives evaluated at the expansion point have been denoted by 
\mbox{$\Gamma^{(n)}(x_1,\ldots,x_n)=\left[\frac{\delta^nF}{\delta\eta
(x_1)\ldots\delta\eta(x_n)}\right]_{\eta\equiv0}$}, \mbox{$\forall n \geq
1$}, and $\Gamma^{(0)} \equiv F[b]$. Since those functional derivatives are
evaluated at $\psi = b$, they must exhibit the same symmetries that leave
the geometry of that system (two concentric circular cylinders) invariant.
The symmetry group contains translations in $z$ and rotations in $\varphi$.
Therefore, we conclude that $\Gamma^{(1)}$ can only be a constant function,
and that $\Gamma^{(2)}(x_1,x_2)$ may only depend on the difference
$x_1-x_2$.  Furthermore, for $n > 2$, one can show that $\Gamma^{(n)}$ may
be written in terms of just $n-1$ independent variables, for instance,
$(x_1-x_2,\, x_2-x_3,\,\ldots,\,
x_{n-2}-x_{n-1},\,x_1+\ldots+x_{n-1}-(n-1)x_n)$.

To proceed, as in the case of Cartesian coordinates, we assume that the
radius $b$ has been chosen in such a way that $\int_x \eta(x)=0$; with this
choice,  the $n=1$ term vanishes. Thus, introducing the Fourier transform
of $\eta$:
\begin{equation}
\eta(\varphi,z) \,=\, \frac{1}{2\pi}\sum_{n=-\infty}^\infty
\int_k \widetilde{\eta}(k,n)e^{ikz}e^{in\varphi},
\end{equation}
with $\int_k \equiv \int_{-\infty}^\infty\frac{\mathrm{d}k}{2\pi}$, we see that:
\begin{equation*}
F[\psi] \;=\; F[b] \,+\, \sum_{n\ge 2}\frac{1}{(2\pi)^{n}}
\sum_{m_1,...,m_n}\int_{k_1,...,k_n}
h^{(n)}(k_1,\ldots,k_{n},m_1,\ldots,m_{n})
\end{equation*}
\begin{equation}\label{3FPsi}
\times
 \; \widetilde{\eta}(k_1,m_1)\ldots\widetilde{\eta}(k_n,m_n)\, 
\delta(k_1+\ldots+k_n)\,\delta(m_1+\ldots+m_n),
\end{equation}
where $h^{(n)}(k_1,\ldots,k_{n},m_1,\ldots,m_{n})$ are the symmetrized form factors.

Based on the previous expressions, we now deal with the zeroth and second
order terms in the DE (the first order one vanishes by the proper choice of
$b$).

\subsection{Zeroth order in derivatives}\label{S:3Orden0}
When $\eta$ becomes sufficiently smooth, $\widetilde{\eta}(k,m)$ is concentrated 
around zero momentum, namely, $(k=0,m=0)$. 
The leading term in this expansion amounts to keeping just that component, namely, to replacing
in (\ref{3FPsi}) the form factors by their zero-momentum limits. 

Hence,
\begin{equation*}
F[\psi]\simeq F_0[\psi]\equiv F[b]+\sum_{n\ge 
2}h^{(n)}(0,\ldots,0)\,\int_{x_1,...,x_{n}}\frac{1}{(2\pi)^{n}}
\sum_{m_1,...,m_n}\int_{k_1,...,k_n}
\eta(x_1)\ldots\eta(x_n)\,
\end{equation*}
\begin{equation}
\times e^{-i\sum_{j=1}^n k_jz_j}e^{-i\sum_{j=1}^n m_j\varphi_j}\, 
\delta(k_1+\ldots+k_n)\,\delta(m_1+\ldots+m_n).
\end{equation}
By taking into account the presence of the delta functions, we can perform 
both the integral over $k_n$ and the sum over $m_n$, obtaining:
\begin{equation}\label{3F0}
F_0[\psi]=F[b]+\int_{x}\,\sum_{n\ge 
2}\frac{h^{(n)}(0,\ldots,0)}{(2\pi)^2}\,\eta(x)^n \;.
\end{equation}
Let us now deal with the evaluation of the sum 
\begin{equation}
\sum_{n\ge 2}h^{(n)}(0,\ldots,0) \left(\eta(x)\right)^n/(2\pi)^2
\end{equation}
as a function of $x$, considering a constant $\eta \equiv \eta_0$,
for which we get
\begin{equation}
\mathcal{F}_0(b+\eta_0)=\mathcal{F}_0(b)+\frac{1}{b}\,\sum_{n\ge 
2}\frac{h^{(n)}(0,\ldots,0)}{(2\pi)^2}\,\eta_0^n,
\end{equation}
where $\mathcal{F}_0(b)$ denotes the  function:
\begin{equation}
\mathcal{F}_0(b)=\lim_{L \to \infty}\frac{F[b]}{S_{b,L}} \;,
\end{equation}
where $S_{b,L}$ denotes the total area of the cylinder $r=b$ and length $L$.

Hence, we extract the relation:
\begin{equation}
\frac{1}{b}\,\sum_{n\ge 
2}\frac{h^{(n)}(0,\ldots,0)}{(2\pi)^2}\,\eta(x)^n=\mathcal{F}
_0(b+\eta(x))-\mathcal{F}_0(b)\;.
\end{equation}
Using the expression above in (\ref{3F0}), we see that:
\begin{equation}
F_0[\psi]=\int_x b\,\mathcal{F}_0(b+\eta(x))=\int_{-\pi}^\pi
\mathrm{d}\varphi\int_{-\infty}^\infty\mathrm{d}z\,b\,\mathcal{F}
_0(b+\eta(\varphi,z)) \;.
\end{equation}

Note that the expression above is quite different to 
the would-be zeroth order result for the DE based on planar surfaces. 
In fact, that would mean to integrate the energy per unit 
area for planes, $\mathcal{F}_0^{P}$, over a planar surface $L$. Indeed, 
for  a physical problem described by two surfaces defined in cylindrical coordinates by
$\rho=r_1$ and $\rho=r_2+\eta(z,\varphi)$, this planar PFA yields
\begin{equation}
F_0^{P}=\int_S \mathcal{F}_0^{P}(r_2+\eta(z,\varphi)-r_1),
\end{equation}
where $S$ is some intermediate surface, and, clearly, the result will in general 
depend on the choice of the surface $S$. 
This is not so for $F_0$. 

The reason for the difference between the two approaches is of course the
fact that the density ${\mathcal F}_0$ generally depends on both $b$ and
$a$ independently, not just on their difference like it necessarily
happens for ${\mathcal F}_0^P$. As a simple example of this situation, we
recall the case of the electrostatic interaction between two conducting
surfaces, held at a constant potential difference, where ${\mathcal F}_0$
is a function of $\log(b/a)$.  
\subsection{Higher orders}\label{S:3Orden2}
To obtain higher order terms in the expansion in derivatives, we need the
corresponding terms in the  Taylor expansion of the momentum space form
factors at zero momentum. Assuming the expansion is well defined,
\begin{equation*}
h^{(n)}(k_1,\ldots,k_n,m_1,\ldots,m_n)=h^{(n)}(0,\ldots,0)+A^{(n)i}k_i+B^{
(n)i}m_i
\end{equation*}
\begin{equation}\label{3hn2}
+C^{(n)ij}k_ik_j+D^{(n)ij}m_im_j+E^{(n)ij}m_ik_j+ \ldots .
\end{equation}
Besides, the variables $m_i$ are integers. However, the analyticity of  
$h^{(n)}$ for $m_i$ regarded as real variables is a sufficient condition 
for the validity and unicity of the expansion (\ref{3hn2}).

To study the consequences of (\ref{3hn2}), we may first note that the
coefficients $C^{(n)ij}$ and $D^{(n)ij}$ can be regarded as invariant
under the exchange of arbitrary $i$ and $j$, since they are multiplied
respectively by $k_ik_j$ and $m_im_j$. On the other hand, 
$h^{(n)}{(\{k_i\},\{m_i\})}$ must be invariant under the exchange of any
two pairs $(k_l,m_l)$ and $(k_s,m_s)$. Therefore, using (\ref{3hn2}), 
we can calculate the difference between $h^{(n)}$ and the same factor when two
such pairs are exchanged, up to order $2$ in $\{m_i\},\,\{k_i\}$. This
gives the following relation for every $l,s$:
$$
0=(A^l-A^s)(k_l-k_s)+(B^l-B^s)(m_l-m_s)+
(C^{ll}-C^{ss})(k_lk_l-k_sk_s) 
$$
$$
+2\sum_{i\neq s,l}k_i(k_l-k_s)(C^{il}-C^{is})+2\sum_{i\neq 
s,l}m_i(m_l-m_s)(D^{il}-D^{is})
$$
$$
+ \sum_{j\neq s,l}k_j(m_l-m_s)(E^{lj}-E^{sj})
+\sum_{i\neq 
s,l} \Big[ m_i(k_l-k_s)(E^{il}-E^{is})
$$
\begin{equation}
+(E^{ls}-E^{sl})(m_lk_s-m_sk_l)+(E^{ll}-E^{
ss})(m_lk_l-m_sk_s) \Big]\;,
\end{equation}
where the indices $(n)$ have been omitted. Using this equation, we can 
obtain useful relations involving the coefficients $A^i$, $B^i$, $C^{ij}$, 
$D^{ij}$, and $E^{ij}$. For instance, setting $k_i=0$ except for $k_l$ and 
$k_s$, and every $m_i$ equal to $0$, we obtain:
\begin{equation}
0=(A^l-A^s)(k_l-k_s)+(C^{ll}-C^{ss})(k_lk_l-k_sk_s),
\end{equation}
from where $A^l=A^s$ and $C^{ll}=C^{ss}$ for every $l,s$, since otherwise 
the functions $(k_l-k_s)$ and $(k_lk_l-k_sk_s)$ would be linearly 
dependent. In a similar fashion, another set of relations may be obtained:
\begin{equation}\label{3RelacionesCoefs1}
A^l=A^s\,,\,B^l=B^s\,,\,C^{ll}=C^{ss}\,,\,
D^{ll}=D^{ss}\,,\,E^{ll}=E^{ss}\,,\,E^{ls}=E^{sl}\,\,\,\forall\, l,s
\end{equation}
\begin{equation}\label{3RelacionesCoefs2}
C^{rl}=C^{rs}\,,\,D^{rl}=D^{rs}\,,\,E^{rl}=E^{rs}\,\,\,
\forall\, l,s,r/l\neq s,\,s\neq r,\,l\neq r.
\end{equation}
Using this result, Eq.~(\ref{3hn2}) may be rendered as
\begin{equation*}
h^{(n)}(k_1,\ldots,k_n,m_1,\ldots,m_n)=h^{(n)}(0,\ldots,0)+
A^{(n)}\sum_{i}k_i+B^{(n)}\sum_{i}m_i
\end{equation*}
\begin{equation*}
+\left[
C_1^{(n)}\sum_{i}k_ik_i
+C_2^{(n)}\sum_{i>j}k_ik_j+
D_1^{(n)}\sum_{i}m_im_i+D_2^{(n)}\sum_{i>j}m_im_j
\right.
\end{equation*}
\begin{equation}\label{3Ec_corchetes}
\left.+E_1^{(n)}\sum_{i}m_ik_i+E_2^{(n)}\sum_{i>j}
\frac{m_ik_j+m_jk_i}{2}
\right]+\ldots
\end{equation}
Replacing this in equation (\ref{3FPsi}), we can see that the term
$h^{(n)}(0,\ldots,0)$ gives rise to the functional $F_0$ already calculated
in (\ref{S:3Orden0}). On the other hand, the linear terms are multiplied by
$\delta(k_1+\ldots+ k_n)\delta(m_1+\ldots m_n)$, and therefore their
contribution vanishes. We are finally left with the order-$2$ terms,
highlighted between brackets in (\ref{3Ec_corchetes}). Performing analogous
steps to those followed in the previous section, we obtain that these
produce a contribution $F_2[\psi]$ given by
\begin{equation*}
F_2[\psi]=-\frac{1}{(2\pi)^2}\sum_{n\ge 2}\int_x\left[
C_1^{(n)}n\,\eta(x)^{n-1}\partial_z^2\eta(x)+
C_2^{(n)}n(n-1)\,\eta(x)^{n-2}\left(\partial_z\eta(x)\right)^2
\right.
\end{equation*}
\begin{equation*}
+D_1^{(n)}n\,\eta(x)^{n-1}\partial_\varphi^2\eta(x)+D_2^{(n)}n(n-1)\,
\eta(x)^{n-2}\left(\partial_\varphi\eta(x)\right)^2
\end{equation*}
\begin{equation}
\left.
+E_1^{(n)}n\,\eta(x)^{n-1}\partial_z\partial_\varphi\eta(x)+
E_2^{(n)}n(n-1)\,\eta(x)^{n-2}\left(\partial_z\eta(x)\right)
\left(\partial_\varphi\eta(x)\right)
\right].
\end{equation}
Now we can perform an integration by parts of the terms that are
proportional to $C^{(n)}_1$, $D^{(n)}_1$ and $E^{(n)}_1$. For instance, for
those proportional to $C^{(n)}_1$, we may do what follows:
\begin{equation*}
\int_{-\pi}^\pi\mathrm{d}\varphi\int_{-\infty}^\infty
\mathrm{d}z\,
\eta(x)^{n-1}\partial_z^2\eta(x)=
-(n-1)\int_{-\pi}^\pi\mathrm{d}\varphi\int_{-\infty}^\infty
\mathrm{d}z\,
\eta(x)^{n-2}\left(\partial_z\eta(x)\right)^2
\end{equation*}
\begin{equation}
+\int_{-\pi}^\pi\mathrm{d}\varphi
\left[\eta(x)^{n-1}\partial_z\eta(x)\right]_{-\infty}^\infty
=-(n-1)\int_{-\pi}^\pi\mathrm{d}\varphi\int_{-\infty}^\infty
\mathrm{d}z\,
\eta(x)^{n-2}\left(\partial_z\eta(x)\right)^2,
\end{equation}
if $\eta(\varphi,z)$ or $\partial_z\eta(\varphi,z)$ vanish as $\vert
z\vert\rightarrow\infty$. A similar procedure can be done for the term
proportional to $D_1^{(n)}$, in the case that $\eta(\varphi,z)$ and
$\partial_\varphi\eta(\varphi,z)$ are periodic functions in $\varphi$,
with period $2\pi$. Doing the same with the term proportional to
$E_1^{(n)}$, we arrive to the desired expression for the order 2 of
$F[\psi]$ in derivatives of $\eta$:
\begin{equation}
F_2[\psi]=\int_x\bigg\{
Z_1(\psi(x))\left(\partial_z\psi\right)^2+
Z_2(\psi(x))\left(\partial_\varphi\psi\right)^2+
Z_{12}(\psi(x))\left(\partial_\varphi\psi\right)
\left(\partial_z\psi\right)
\bigg\},
\end{equation}
where the functions $Z_1(b+d)$, $Z_2(b+d)$ and $Z_{12}(b+d)$ are defined as:
\begin{equation}
Z_1(b+d)=\sum_{n\ge 
2}\frac{n(n-1)}{(2\pi)^2}\left[C_1^{(n)}-C_2^{(n)}\right]d^{n-2},
\end{equation}
\begin{equation}
Z_2(b+d)=\sum_{n\ge 
2}\frac{n(n-1)}{(2\pi)^2}\left[D_1^{(n)}-D_2^{(n)}\right]d^{n-2},
\end{equation}
\begin{equation}
Z_{12}(b+d)=\sum_{n\ge 
2}\frac{n(n-1)}{(2\pi)^2}\left[E_1^{(n)}-E_2^{(n)}\right]d^{n-2}.
\end{equation}
To calculate these functions in a simpler way, we may evaluate them for 
$d=0$, which gives us their value in $b$:
\begin{equation}
Z_1(b)=\frac{1}{2\pi^2}\left[C_1^{(2)}(b)-C_2^{(2)}(b)\right],
\end{equation}
\begin{equation}
Z_2(b)=\frac{1}{2\pi^2}\left[D_1^{(2)}(b)-D_2^{(2)}(b)\right],
\end{equation}
\begin{equation}
Z_{12}(b)=\frac{1}{2\pi^2}\left[E_1^{(2)}(b)-E_2^{(2)}(b)\right].
\end{equation}
Finally, to obtain the order two of $F[\psi]$, it is enough to change $b$
for $\psi(x)$ in the argument of these functions, and to replace them in
equation (\ref{3DE_Cilindricas}). 

Thus, the second order DE is
\begin{align}\label{3DE_Cilindricas}
	F[\psi] &=\int_x b\,\mathcal{F}_0(b+\eta(x)) \nonumber\\
&+\int_x\bigg\{
Z_1(\psi(x))\left(\partial_z\psi\right)^2+
Z_2(\psi(x))\left(\partial_\varphi\psi\right)^2+
Z_{12}(\psi(x))\left(\partial_\varphi\psi\right)
\left(\partial_z\psi\right)
\bigg\}\;.
\end{align}
We recall that we have assumed $F[\psi]$ to be analytic (as a functional) in a
neighbourhood of $\psi\equiv a$ as well as the form factors $h^{(n)}$ at zero 
momenta. On the other hand, $\eta$ and $\partial_z\eta$ must tend to $0$ as
$\vert z \vert\rightarrow\infty$. Finally, $\psi$ and
$\partial_\varphi\psi$ are periodic functions of $\varphi$, with 
period $2\pi$. Except for the last condition, the other are
equivalent to those required to apply the DE in Cartesian
coordinates~\cite{Fosco:2014rfa}. In addition, the method provides a tool to calculate the following
orders, namely, by including higher order products of  $k_i$ and $m_i$ in
the expansion (\ref{3hn2}).  

Note that, from (\ref{3DE_Cilindricas}), the order 2 we obtain
is not proportional to the square of the gradient of $\psi$, as it happened in Cartesian
coordinates. This is because, when considering the interaction energy
between two planes $x_3=0$ and $x_3=\psi(x_1,x_2)$ in an isotropic space,
the functional $F[\psi]$ must be invariant under rotations in the argument
of $\psi$, this is, if $(x_1,x_2)^T$ is replaced by $R.(x_1,x_2)^T$, with
$R\in SO(2)$. This symmetry is however lost when considering functions
$\psi(\varphi,z)$, which justifies the mixed term in
(\ref{3DE_Cilindricas}).
In the presence of extra symmetries (which may even be discrete), one could
of course say more about the vanishing of one or more terms in DE. 


\end{document}